\documentclass[twocolumn,showpacs,preprintnumbers,amsmath,amssymb]{revtex4}
\usepackage{graphicx}
\usepackage{dcolumn}
\usepackage{bm}
\begin{document}
\title{Failure properties of loaded fiber bundles having a lower cutoff
in fiber threshold distribution}
\author{Srutarshi\ Pradhan}
\email{pradhan.srutarshi@phys.ntnu.no}
\author{Alex Hansen}
\email{alex.hansen@phys.ntnu.no}
\affiliation{Department of Physics, Norwegian University of Science and
Technology, N--7491 Trondheim, Norway}
\begin{abstract}
\noindent Presence of lower cutoff in fiber threshold distribution
may affect the failure properties of a bundle of fibers subjected
to external load. We investigate this possibility --both in a equal
load sharing (ELS) model and in local load sharing (LLS) one using
analytic as well as numerical methods. In ELS model, the critical
strength gets modified and beyond a certain lower cutoff level the
whole bundle fails instantly (brittle failure) after the first fiber
ruptures. Although the dynamic exponents for the order parameter,
susceptibility and relaxation time remain unchanged, the avalanche
size distribution shows a gradual deviation from the mean field power
law. A similar `instant failure' situation occurs in LLS model at
a lower cutoff level which reduces to that of in the equivalent ELS
model at higher (high enough) dimensions. Also, the system size variation
of bundle's strength and the avalanche statistics show strong dependence
on the lower cutoff level. 
\end{abstract}
\maketitle

\section{Introduction}

Critical behavior of the fracture-failure phenomena in disordered
materials has attracted wide interest these days \cite{books}. Among
several model studies, the fiber bundle models (FBM) capture almost
correctly the collective static and dynamics of fracture-failure in
loaded materials. The two different versions of FBM have been studied
much. The equal load sharing (ELS) model \cite{Peirce,Dan45} considers
democratic (equal) sharing of applied load on the bundle, whereas
in local load sharing (LLS) model \cite{LLS,LLS-th}, only the nearest
neighbors support the terminal load (stress) of a failed fiber. Experimentally
it has been observed \cite{AE-97,AE-98,AE_94} that disordered material
under increasing load shows well defined power laws in terms of acoustic
emissions prior to the global rupture. Such a power law in burst avalanches
has been achieved analytically (and verified through simulations)
by Hemmer and Hansen \cite{HH92,Kloster} in ELS model. It is also
known since several decades that the static ELS model has a critical
point \cite{Phase T,D Sornet,pach-00,RS99,DD-97}, i.e., at a critical
strength ($\sigma_{c}$) the bundle shows a phase transition from
a state of partial failure (for $\sigma\leq\sigma_{c}$) to the state
of total failure (for $\sigma>\sigma_{c}$). This failure dynamics
has been solved analytically \cite{SB01,SBP02} which explores the
critical behavior through the power law variation of order parameter,
susceptibility and relaxation time. Also the mean-filed universality
of ELS model has been established recently \cite{PSB03}. The extensive
studies on LLS model \cite{LLS,LLS-th} suggest that the strength
goes to `zero' value as the bundle size approaches infinity \cite{Smith-80,Pacheco,SB-mod},
--thus excludes the possibility of any critical behavior. Another
important observation is that no universal power law asymptotics exists
for the avalanche statistics \cite{Kloster} in LLS model. Attempts
have also been made to study both, the ELS and LLS models in a single
framework introducing adjustable load sharing parameter \cite{lamda model,variable range,SP-N1-2004}
and a crossover from mean-field (ELS) to short-range (LLS) behavior
has been reported. 

So far in the FBM studies, people mainly considered different fiber
threshold distributions starting from zero threshold. However, in
reality every element (fiber) should have a finite (non-zero) strength
threshold due to the cohesive force among the constituting molecules.
Therefore the idea of a lower cutoff in fiber threshold distribution
would be most welcome. Andersen et al \cite{DD-97} considered first
a lower cutoff in fiber threshold distribution and established the
`tricritical behavior' in the mean-field (ELS) fiber bundle model.
Such distributions with lower cutoff have also been considered to
study the nonlinear response in ELS mode \cite{SBP02} and to establish
the universal behavior \cite{PSB03} of ELS failure dynamics. The
exclusion of weaker fibers not only enhances the ultimate strength
of the bundle, it can affect the failure properties of the bundle.
To investigate such possibilities --in the present work we consider
ELS and LLS models and proceed through analytic as well as numerical
methods. 

We organize this report as follows: After a brief introduction (Section
I) we study the effect of lower cutoff in ELS model (Section II) and
in LLS model (Section III). The importance of such study and the physical
significance of the observed results are discussed in the conclusion
(Section IV). In the Appendix we apply our analytic formulations in
two different situations of fiber threshold distribution.

\section{`ELS' model}

\subsection{Solutions of the recursive dynamics for equal load increment}

\noindent We consider a fiber bundle model having $N$ parallel fibers
subjected to an external load or stress (load per fiber). The threshold
strength of each fiber is determined by the stress value ($\sigma_{th}$)
it can bear, and beyond which it fails. We consider fiber threshold
distribution to have a lower cutoff ($\sigma_{L}$), i.e., a randomly
distributed normalized density $\rho(\sigma_{th})$ has been chosen
within the interval $\sigma_{L}$ and $1$ such that \begin{equation}
\int_{\sigma_{L}}^{1}\rho(\sigma_{th})d\sigma_{th}=1.\label{norm}\end{equation}

\noindent We follow step-wise equal load increment \cite{SB01,SBP02}
till the total failure of the bundle. The breaking dynamics starts
when an initial stress $\sigma$ ($>\sigma_{L}$) is applied on the
bundle. Fibers having strength less than $\sigma$ fail instantly
reducing the number of intact fibers and these fibers have to bear
the applied load (ELS rule). Hence the effective stress (on intact
fibers) increases and this compels some more fibers to break. These
two sequential operations, the stress redistribution and further breaking
of fibers continue till an equilibrium is reached, where either the
surviving fibers are strong enough to bear the applied load or all
fibers fail. 

The breaking dynamics can be represented by a recursion relation \cite{SB01,SBP02,PSB03}
in discrete time steps: 

\noindent \begin{equation}
U_{t+1}=1-P(\sigma/U_{t});U_{0}=1;\label{els-1}\end{equation}
where, $U_{t}$ is the fraction of total fibers that survive after
time step $t$ and $P(\sigma_{t})$ is the cumulative distribution
of corresponding density $\rho(\sigma_{th})$, \begin{equation}
P(\sigma_{t})=\int_{\sigma_{L}}^{\sigma_{t}}\rho(\sigma_{th})d\sigma_{th}.\label{els-2}\end{equation}
The time step indicates the number of stress redistribution at a fixed
applied load. 

At the equilibrium or steady state we get $U_{t+1}=U_{t}\equiv U^{*}$.
This is a fixed point of the recursive dynamics and eqn. (\ref{els-1})
can be solved at the fixed point for some particular strength distribution. 

\includegraphics[%
  width=2.5in,
  height=2in]{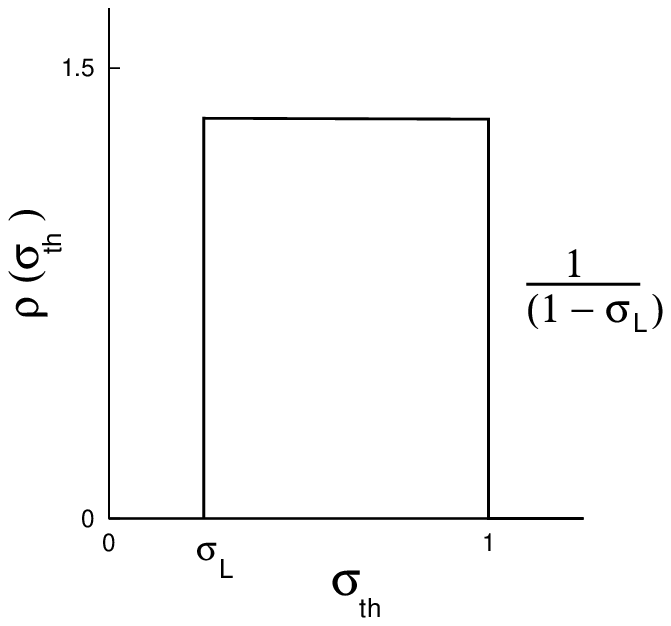}

\vskip .1in

\textbf{\footnotesize Fig. 1:} {\footnotesize The uniform density
of fiber strength having a lower cutoff ($\sigma_{L}$) .}{\footnotesize \par}

\vskip .1in

We choose the uniform density of fiber strength threshold having a
lower cutoff (Fig. 1) to solve the recursive failure dynamics of ELS
model. Thus $\rho(\sigma_{th})$ has the form: \begin{equation}
\rho(\sigma_{th})=\frac{1}{1-\sigma_{L}},\sigma_{L}<\sigma_{th}\leq1.\label{uni-dist}\end{equation}
The cumulative distribution becomes \begin{equation}
P(\sigma_{t})=\int_{\sigma_{L}}^{\sigma_{t}}\rho(\sigma_{th})d\sigma_{th}=\frac{(\sigma_{t}-\sigma_{L})}{(1-\sigma_{L})}.\label{els-3}\end{equation}
Therefore $U_{t}$ follows a simple recursion relation: \begin{equation}
U_{t+1}=\frac{1}{1-\sigma_{L}}\left[1-\frac{\sigma}{U_{t}}\right],\label{els-4}\end{equation}

\noindent which has fixed points \cite{SBP02,PSB03}: \begin{equation}
U^{*}(\sigma)=\frac{1}{2(1-\sigma_{L})}\left[1\pm\left(1-\frac{\sigma}{\sigma_{c}}\right)^{1/2}\right];\label{els-5}\end{equation}

\noindent where

\begin{equation}
\sigma_{c}=\frac{1}{4(1-\sigma_{L})}.\label{sigmac}\end{equation}

\noindent Beyond this critical strength ($\sigma_{c}$) the whole
bundle fails instantly. The solution with ($+$) sign is the stable
one, whereas the one with ($-)$ sign gives unstable solution \cite{SBP02,PSB03}. 

\vskip.1in

\includegraphics[%
  width=1.5in,
  height=1.7in]{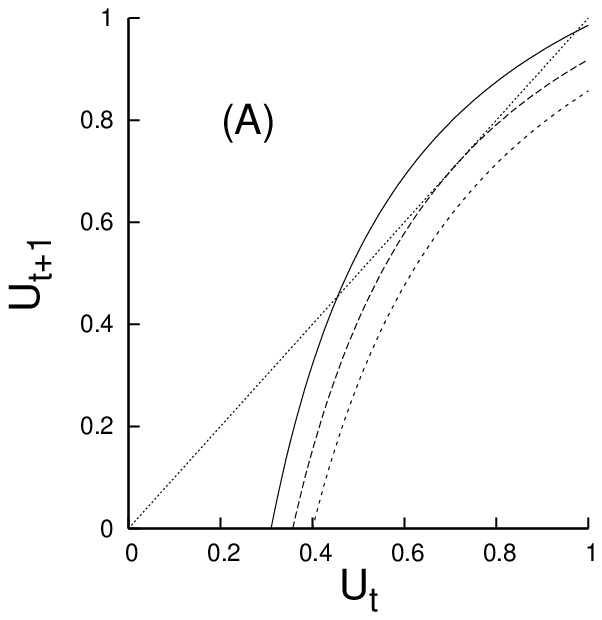}\includegraphics[%
  width=1.5in,
  height=1.7in]{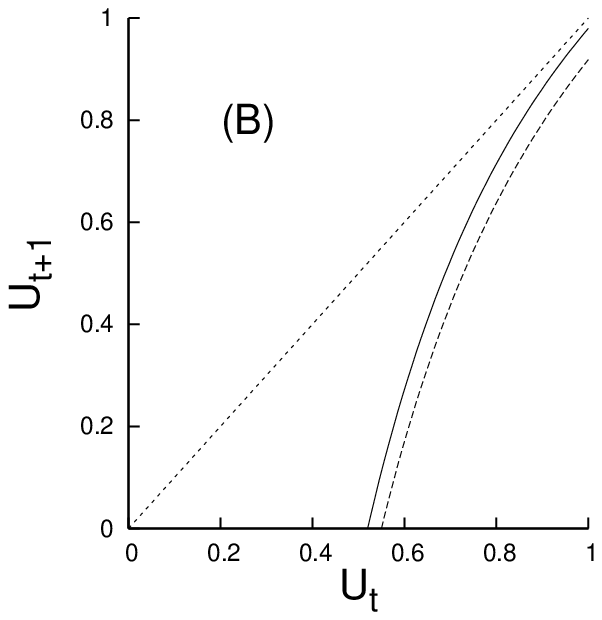}

\vskip.1in

\textbf{\footnotesize Fig. 2:} {\footnotesize The graphical solutions
of eqn. (\ref{els-4}) : Straight lines represent the fixed points.
In (A) $\sigma_{L}=0.3$, $\sigma_{c}=0.357$, solid curve ($\sigma<\sigma_{c}$),
dashed line ($\sigma=\sigma_{c}$), dotted line ($\sigma>\sigma_{c}$).
In (B) $\sigma_{L}=0.51$, solid curve ($\sigma=0.52.$), dashed line
($\sigma=0.55$).}{\footnotesize \par}

\vskip.1in

It is obvious that the critical strength ($\sigma_{c}$) cannot be
less than $\sigma_{L}$. \emph{As $\sigma_{c}$ is a critical point
(see \cite{SB01,SBP02,PSB03}), there should be some critical exponents
associated to $\sigma_{c}$. If none of the fibers fail ($\sigma<\sigma_{L}$),
we cannot define order parameter, susceptibility, relaxation time
etc. which show critical behavior of the failure dynamics. Therefore,
at $\sigma_{c}$ the bundle should be in a partially broken (stable)
state.} Putting the condition $\sigma_{c}\geq\sigma_{L}$ in eqn.
(\ref{sigmac}), we get the upper bound of lower cutoff: $\sigma_{L}<1/2$.
We can verify that for $\sigma_{L}>1/2$ the recursion (eqn. \ref{els-4})
does not give a stable fixed point except $U^{*}=0$. Also, putting
$\sigma_{L}>1/2$ in fixed point solution (eqn. \ref{els-5}), we
get $U^{*}>1$, which is unrealistic. Therefore the critical strength
of ELS model is bounded by an upper limit: $\sigma_{c}\leq1/2$. We
present graphical solutions (Fig. 2) of the recursion relation (eqn.
\ref{els-4}) for $\sigma_{L}<1/2$ (A) and $\sigma_{L}>1/2$ (B).
Clearly, we cannot get a fixed point ($U_{t+1}=U_{t}$) in (B).

From the solution (eqn. \ref{els-5}) we can obtain the order parameter
($O$), susceptibility ($\chi$) and relaxation time ($\tau$) \cite{SB01,SBP02,PSB03}
of the failure process: \begin{equation}
O=U^{*}(\sigma)-U^{*}(\sigma_{c})\sim(\sigma_{c}-\sigma)^{-\alpha};\alpha=\frac{1}{2}\label{alpha}\end{equation}
\begin{equation}
\chi=\left|\frac{dU^{*}(\sigma)}{d\sigma}\right|\sim(\sigma_{c}-\sigma)^{-\beta};\beta=\frac{1}{2}\label{beta}\end{equation}

\begin{equation}
\tau\sim(\sigma_{c}-\sigma)^{-\theta};\theta=\frac{1}{2}.\label{theta}\end{equation}

\noindent Therefore the variation of order parameter, susceptibility
and relaxation time remain unaffected by the presence of lower cutoff.

\subsection{The `instant failure' situation in weakest fiber breaking approach}

Now, we follow the weakest fiber breaking approach \cite{Dan45,HH92}:
The applied load is tuned in such a way that only the weakest fiber
(among the intact fibers) will fail after each step of loading. We
first find out the extreme condition when the whole bundle fails instantly
after the first fiber ruptures. As the strength thresholds of $N$
fibers are uniformly distributed within $\sigma_{L}$ and $1$, the
weakest fiber fails at a stress $\sigma_{L}$ (for large $N$). After
this single fiber failure, the load will be redistributed within intact
fibers resulting a global stress $\sigma_{f}=N\sigma_{L}/(N-1)$.
Now, the number of intact fibers having strength threshold below $\sigma_{f}$
is 

\begin{equation}
NP(\sigma_{f})=N\int_{\sigma_{L}}^{\sigma_{f}}\rho(\sigma_{th})d\sigma_{th}=\frac{N(\sigma_{f}-\sigma_{L})}{(1-\sigma_{L})}.\label{Ad-1}\end{equation}

The stress redistribution can break at least another fiber if $NP(\sigma_{f})\geq1$
and this `second' failure will trigger another failure and so on.
Thus the successive breaking of fibers cannot be stopped till the
complete collapse of the bundle. Clearly, there cannot be any fixed
point (critical point) for such `instant failure' situation. Putting
the value of $\sigma_{f}$ we get 

\begin{equation}
\frac{N(\frac{N\sigma_{L}}{N-1}-\sigma_{L})}{(1-\sigma_{L})}\geq1;\label{Ad-2}\end{equation}

\noindent which gives \begin{equation}
\sigma_{L}\geq\frac{(N-1)}{(2N-1)}.\label{Ad-3}\end{equation}
 For large $N$ limit, the above condition can be written as $\sigma_{L}\geq1/2$.
Therefore, the condition to get a fixed point in the failure process
is $\sigma_{L}<1/2$. 

\vskip.1in

\includegraphics[%
  width=2.5in,
  height=2in]{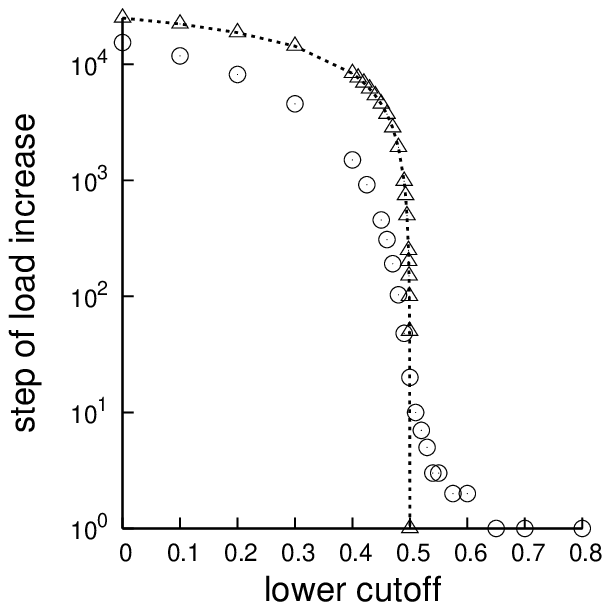}

\vskip.1in

\textbf{\footnotesize Fig. 3:} {\footnotesize The total step of load
increase (till final failure) is plotted against $\sigma_{L}$ for
a ELS model having $50000$ fibers. The dotted line represents the
analytic form (eqn. \ref{step no}), triangles are the simulated data
for a strictly uniform strength distribution and the circles represent
the data (averages are taken for $5000$ samples) for a uniform on
average distribution. }{\footnotesize \par}

\vskip.1in

We can also calculate how many steps are required to attain the final
catastrophic failure for $\sigma_{L}<1/2$. Let us assume that we
have to increase the external load $x$ times before the final failure.
At each step of such load increment only one fiber fails. Then after
$x$ step the following condition should be fulfilled to have a catastrophic
failure:

\begin{equation}
N\int_{\sigma_{i}}^{\sigma_{i}[1+1/(N-x)]}\rho(\sigma_{th})d\sigma_{th}\geq1.\label{uni-step-1}\end{equation}

\noindent where \begin{equation}
\sigma_{i}=\sigma_{L}+\frac{x(1-\sigma_{L})}{N}\label{uni-step-2}\end{equation}

\noindent The solution gives \begin{equation}
x=\frac{N}{2}\left(1-\frac{\sigma_{L}}{1-\sigma_{L}}\right).\label{step no}\end{equation}

\noindent The above equation suggests that at $\sigma_{L}=1/2$, $x=0$.
But in reality we have to put the external load once to break the
weakest fiber of the bundle. Therefore, $x=1$ for $\sigma_{L}\geq1/2$
(Fig. 3). To check the validity of the above calculation we take `strictly
unifrom' and uniform on average distributions of fiber strength. In
our `strictly unifrom' distribution the strength of the $k$-th fiber
(among $N$ fibers) is $\sigma_{L}+(1-\sigma_{L})k/N$. We can see
in Fig. 3 that the `strictly uniform distribution' exactly obeys the
analytic formula (\ref{step no}) but the uniform on average distribution
shows slight disagreement which comes from the fluctuation in the
distribution function for finite system size. This fluctuation will
disappear at the limit $N\rightarrow\infty$ where we expect perfect
agreement.

\subsection{Avalanche size distribution}

During the failure process `avalanches' of different size appear where
simultanious failure of numbers of fiber is termed as `avalanche'.
To investigate whether the avalanche size distributions depend on
the lower cutoff or not, we go for a numerical study. The result (Fig.
4) demands that for small avalanche sizes the distributions show a
gradual deviation (depend on $\sigma_{L}$) from the mean-field result,
although the big avalanches still follow the mean-field power law
(exponent value $-5/2$) as analytically derived by Hemmer and Hansen
\cite{HH92}. We have checked this result for several system sizes
and the above feature remains invariant. We should mention that according
to eqns. (37-39) of Ref. \cite{HH92}, the analytic treatment is valid
for $\sigma_{L}<1/2$ as $1/2$ is the maximum (upper limit of the
integration in eqn. (37)) of the stress-strain curve. Presence of
$\sigma_{L}$ cuts out the lower part of the stress-strain curve where
the small avalanches are most likely to happen. Therefore, smaller
avalanches get reduced in number with the increase of $\sigma_{L}$
(Fig. 4). At the limit $\sigma_{L}\rightarrow1/2$, the avalanche
distributions seem to follow a new power law with exponent $-3/2$,
which can be explained as: The fluctuation in threshold distribution
gives rise a scenario like the unbiased random walk of the bundle's
strength around the maximum $1/2$, which in turn results exponent
$-3/2$ in avalanche distribution \cite{D Sornet}. 

\vskip.1in

\includegraphics[%
  width=2.5in,
  height=2in]{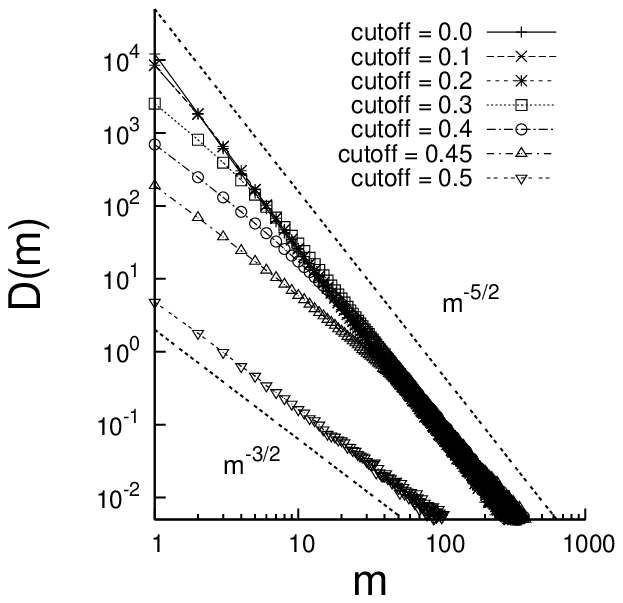}

\vskip.1in

\textbf{\footnotesize Fig. 4:} {\footnotesize The avalanche size distributions
for different values of $\sigma_{L}$: $N=50000$ and averages are
done over $10000$ sample. We have drawn two power laws (dotted lines)
as reference lines to compare our numerical results. }{\footnotesize \par}

\section{`LLS' model}

\subsection{The `instant failure' situation}

Now we consider LLS model with uniform fiber threshold distribution
having a lower cutoff $\sigma_{L}$ (Fig. 1). We shall present a probabilistic
argument to determine the upper limit of $\sigma_{L}$, beyond which
the whole bundle fails at once. Following the weakest fiber breaking
approach the first fiber fails at an applied stress $\sigma_{L}$
(for large $N$). As we are using periodic boundary conditions, the
$n$ nearest neighbors ($n$ is the coordination number) bear the
terminal stress of the failing fiber and their stress value rises
to $\sigma_{f}=\sigma_{L}(1+1/n)$. Now, number of nearest neighbors
(intact) having strength threshold below $\sigma_{f}$ is $(nn)_{fail}=nP(\sigma_{f})$
(see eqn. (\ref{Ad-1})). Putting the value of $P(\sigma_{f})$ and
$\sigma_{f}$ we finally get \begin{equation}
(nn)_{fail}=\frac{(\sigma_{L})}{(1-\sigma_{L})}.\label{eq:lls-prob}\end{equation}
If $(nn)_{fail}\geq1$, then at least another fiber fails and this
is likely to trigger a cascade of failure events resulting complete collapse 
of the bundle. Therefore, to avoid the `instant failure' situation we
must have $(nn)_{fail}<1$, from which we get the upper bound of $\sigma_{L}$: 

\noindent \begin{equation}
\sigma_{L}<\frac{1}{2}.\label{lls-prob2}\end{equation}

\noindent As the above condition does not depend on the coordination
number $n$, at any dimension the whole bundle is likely to collapse
at once for $\sigma_{L}\geq1/2$. It should be mentioned that LLS
model should behave almost like ELS model at the limit of infinite
dimensions and therefore the identical bound (of $\sigma_{L}$) in
both the cases is not surprising (see Appendix). We numerically confirm
(Fig. 5) the above analytic argument (eqn. \ref{lls-prob2}) in one
dimension. When average step value goes below 1.5 line, one step failure 
is the dominating mode then. We can find out the extreme limit of 
$\sigma_{L}$ when all the
nearest neighbors fail after the weakest fiber breakes. Then the LLS 
bundle collapses instantly for sure. Setting $(nn)_{fail}=n$ we get 
the condition  $\sigma_{L}\geq n/(1+n)$, where stress level of all the 
nearest neighbors crosses the upper cutoff 1 of the strength distribution. 
Clearly such failure is very rapid (like a chain reaction) and does not 
depend on the shape of the strength distributions, except for the upper 
cutoff. Also as $n$ increases (ELS limit) - $\sigma_{L}$ for instant failure 
assumes the trivial value 1. Similar sudden failure in FBM 
has been discussed by Moreno
et al \cite{Pacheco -01} in the context of a `one sided load transfer'
model which is different from the true LLS model --we consider here. 

\vskip.1in

\includegraphics[%
  width=2.5in,
  height=2in]{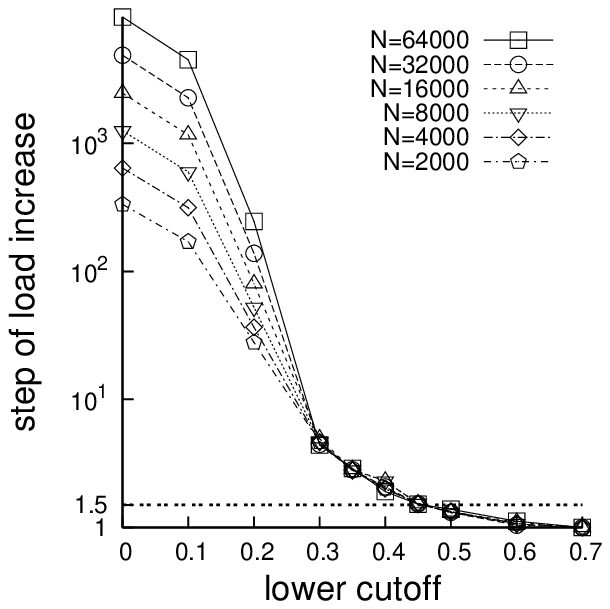}

\vskip.1in

\textbf{\footnotesize Fig. 5:} {\footnotesize Numerical estimate of
the upper bound of $\sigma_{L}$ in LLS model: For $\sigma_{L}\geq0.5$
the average step values go below $1.5$, i.e., the bundle fails at
one step in most of the realisations. }{\footnotesize \par}

\subsection{Strength of the bundle}

The local load sharing (LLS) scheme introduces stress enhancement
around the failed fiber, which accelerate damage evolution. Therefore,
a few isolated cracks can drive the system toward complete failure
through growth and coalescence. The LLS model shows zero strength
(for fiber threshold distributions starting from zero value) at the
limit $N\rightarrow\infty$, following a logarithmic dependence on
the system size ($N$) \cite{Smith-80,Pacheco,SB-mod}. Recently Mahesh
et al \cite{Mahesh} have proposed a probabilistic method of finding
the asymptotic strength of bundles in LLS mode. Now for threshold
distributions having a lower cutoff ($\sigma_{L}$), the ultimate
strength of the bundle cannot be less than $\sigma_{L}$. For such
a uniform distribution (Fig. 1), we perform numerical simulations
to investigate the system size variation of bundle's strength. We
observe (Fig. 6) that as $\sigma_{L}$ increases the quantity (strength-$\sigma_{L}$)
approaches zero following straight lines with $1/N$, but the slope
gradually decreases --which suggests that the system size dependence
of the strength gradually becomes weaker. 

\vskip.1in

\includegraphics[%
  width=2.5in,
  height=2.0in]{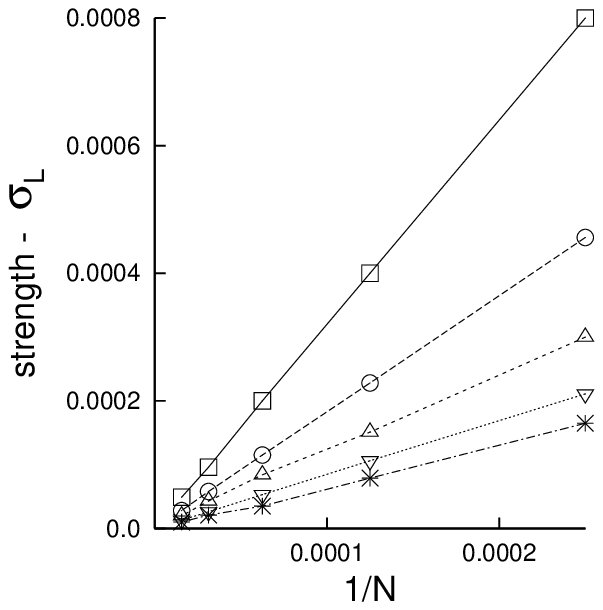}

\vskip.1in

\textbf{\footnotesize Fig. 6:} {\footnotesize The (strength -- $\sigma_{L}$)
is plotted against $1/N$ for different $\sigma_{L}$ values: $0.3$
(square), $0.35$ (circle), $0.4$ (up triangle), $0.45$ (down triangle),
$0.5$ (star)}. {\footnotesize All the straight lines approach $0$
value as $N\rightarrow\infty$. }{\footnotesize \par}
\subsection{Avalanche size distribution}

\includegraphics[%
  width=2.8in,
  height=2.0in]{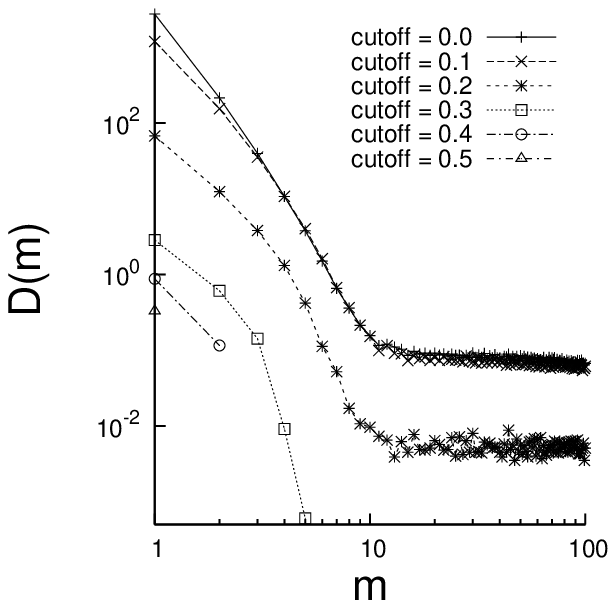} 

\vskip.05in

\textbf{\footnotesize Fig. 7:} {\footnotesize Avalanche size distribution
for several $\sigma_{L}$ in LLS model: $N=20000$, averages are taken
over $10000$ samples. Number of different avalanches
decreases with the increase of $\sigma_{L}$value. }{\footnotesize \par}

\vskip.05in

Due to faster crack-growth, LLS model shows different avalanche statistics
than that of in ELS model. The numerical study of Hansen and Hemmer
\cite{HH92} suggests an apparent power law having exponent $-4.5$
in the avalanche distribution. Later, Kloster et al \cite{Kloster}
have shown analytically that for flat (uniform) distribution LLS model
does not have any power law asymptotics in avalanche statistics. We
numerically study the avalanche distribution in LLS model for different
$\sigma_{L}$ values (Fig. 7). We observe a similar deviation (lowering)
of the distribution function for the smaller avalanche sizes as in
case of ELS model. Also, the number of different avalanches gets reduced
(tail of the distribution disappears) with the increase of $\sigma_{L}$.
This occurs due to damage localization \cite{Kun-00} which ensures
faster collapse of the bundle. In Fig. 7 we can see that for $\sigma_{L}=0.5$,
avalanches of size $1$ is the only possibility before total failure
and their count is always less than $1$, which clearly indicates
the dominance of `instant failure' situation.

\section{Conclusion}

A lower cutoff in fiber threshold distribution excludes the presence
of very weak fibers in a bundle. The weaker fibers mainly reduces
the strength of a bundle. But in practical purpose we always try to
build stronger and stronger materials (ropes, cables etc.) from the
fibrous elements. Therefore this situation (exclusion of weaker fibers)
is very realistic. The failure dynamics of ELS model almost remains
unchanged in the presence of such lower cutoff ($\sigma_{L}$), whereas
the avalanche size distributions show a systematic deviation (for
small avalanches) from the mean field nature. At the limiting point
($\sigma_{L}\rightarrow1/2$), we get a new power law (exponent $-3/2$)
in avalanche distribution which can be explained from random walk
statistics \cite{D Sornet}. In LLS model the avalanche statistics
show drastic change with the increase of $\sigma_{L}$. In both the
models, the lower cutoff becomes bounded by an upper limit ($\sigma_{L}<1/2$)
beyond which the whole bundle fails at once, which has important consequence:
It seems that the bundles show elastic-like response \cite{SBP02}
up to $\sigma_{L}=1/2$, above which they become perfectly brittle.
We observe that the `weakest fiber breaking approach' \cite{Dan45,HH92}
and the `equal load increment approach' \cite{SB01,SBP02} give similar
result (in ELS mode). In the equal load increment method, sometimes
more than one fibers fail at the time of loading and this affects
the whole failure dynamics, whereas the weakest fiber breaking approach
ensures the single fiber (weakest among the intact fibers) failure
at each step of loading. We consider the equal load increment method
to be more practical from the experimental point of view. This approach
helps to construct the recursion relations \cite{SB01,SBP02} which
in turn show critical behavior \cite{SBP02,PSB03} of the failure
process. The `instant failure' situation is not limited to uniform
threshold distribution, rather it is common in any type of distributions
(see Appendix). It seems that the `instant failure' represents the
binary states of the bundle: `intact' (1 or high) and `completely
broken' (0 or low) and here the bundle behaves like a classical `switch'
in response to external load. 

\vskip.3in

\textbf{Acknowledgment}: We like to thank Prof. P. C. Hemmer and Dr.
M. Kloster for useful comments and suggestions. S. P. thanks the Norwegian
Research Council, NFR for the funding through a strategic university
program.

\vskip.5in

\begin{center}\textbf{\LARGE Appendix}\end{center}{\LARGE \par}

\vskip.2in

\textbf{\large Case I : Linearly increasing density of fiber strength}{\large \par}

\includegraphics[%
  width=2.5in,
  height=1.9in]{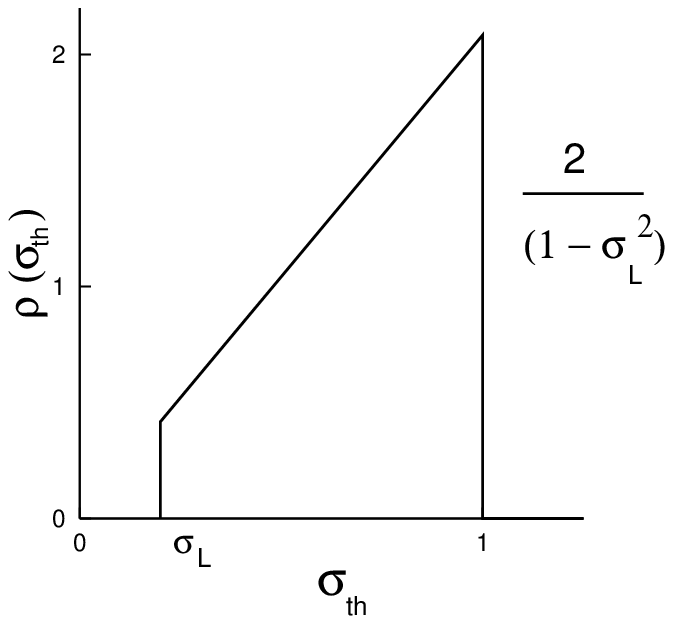}

\vskip.1in

\textbf{\footnotesize Fig. 8:} {\footnotesize The linearly increasing
density of fiber strength having a lower cutoff ($\sigma_{L}$) .}{\footnotesize \par}

\vskip.2in

We consider a bundle of fibers with linearly increasing density of
strength having the normalised form: 

\begin{equation}
\rho(\sigma_{th})=\frac{2\sigma_{th}}{1-\sigma_{L}^{2}},\sigma_{L}\leq\sigma_{th}\leq1.\label{inc-dist}\end{equation}
We want to find out the bound of $\sigma_{L}$ beyond which `instant
failure' occurs in both --ELS and LLS models.

\vskip.1in
\textbf{A. ELS model}
\vskip.1in
Following the weakest fiber breaking approach, the condition for `instant
failure' is:\begin{equation}
N\int_{\sigma_{L}}^{\sigma_{L}[1+1/(N-1)]}\rho(\sigma_{th})d\sigma_{th}\geq1.\label{inc-1-ELS}\end{equation}
which gives

\begin{equation}
\sigma_{L}^{2}\geq\frac{N-1}{3N-1}\label{inc-condition-ELS}\end{equation}
 Therefore, the bound (beyond which one-instant failure will occur)
of the lower cutoff comes to be $\sigma_{L}<1/\sqrt{3}$ for large
$N$ limit. 
\vskip.4in
\textbf{B. LLS model}
\vskip.1in
The condition for `instant failure' for LLS model is: \begin{equation}
n\int_{\sigma_{L}}^{\sigma_{L}[1+1/n]}\rho(\sigma_{th})d\sigma_{th}\geq1.\label{inc-1-LLS}\end{equation}
where $n$ is the coordination number or the number of nearest neighbors.
This gives

\begin{equation}
\sigma_{L}^{2}\geq\frac{1}{\left(3+\frac{1}{n}\right)}.\label{inc-condition-LLS}\end{equation}
Now, as dimension of the system increases $n$ goes towards infinity.
Hence the above condition gives the bound of the lower cutoff as $\sigma_{L}<1/\sqrt{3}$,
which is identical to that of in the equivalent ELS case. 

\vskip.1in

\textbf{\large Case II : Linearly decreasing density of fiber strength}{\large \par}

\includegraphics[%
  width=2.5in,
  height=1.8in]{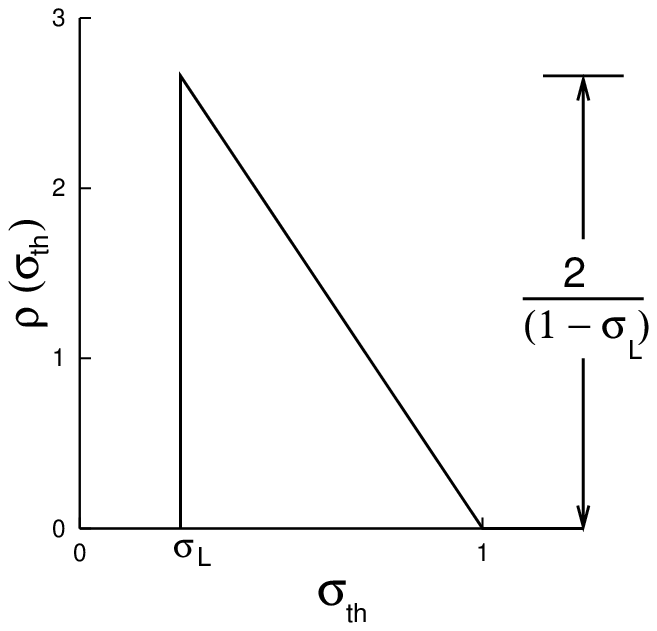}

\vskip.1in

\textbf{\footnotesize Fig. 9:} {\footnotesize The linearly decreasing
density of fiber strength having a lower cutoff ($\sigma_{L}$) .}{\footnotesize \par}

\vskip.1in

Next we consider a bundle of fibers having linearly decreasing density
of strength with the normalised form: 

\begin{equation}
\rho(\sigma_{th})=\frac{2(1-\sigma_{th})}{(1-\sigma_{L})^{2}},\sigma_{L}\leq\sigma_{th}\leq1.\label{dec-dist}\end{equation}

\vskip.1in

\textbf{A. ELS model}

\vskip.1in

Following eqn. ( \ref{inc-1-ELS}), the condition for `instant failure'
is: \begin{equation}
\sigma_{L}\geq\frac{N-1}{3N-1}\label{inc-condition-ELS}\end{equation}
which sets the bound of the lower cutoff to $\sigma_{L}<1/3$. 

\vskip.1in

\textbf{B. LLS model}

\vskip.1in

Following eqn. ( \ref{inc-1-LLS}), the condition for `instant failure'
is: \begin{equation}
\frac{2\sigma_{L}}{(1-\sigma_{L})^{2}}\left(1-\sigma_{L}-\frac{\sigma_{L}}{2n}\right)\geq1.\label{inc-condition-LLS}\end{equation}
 which reduces to $2\sigma_{L}/(1-\sigma_{L})\geq1$ for $n\rightarrow\infty$
and sets the bound of the lower cutoff to $\sigma_{L}<1/3$. Again
this is identical to that of in the ELS case.

\end{document}